\documentclass[amsmath,  showpacs, aps, twocolumn]{revtex4}
\usepackage{amssymb}
\usepackage{epsfig}
\usepackage{natbib}

\begin{document}

\title{First Passage Time Densities in Resonate-and-Fire Models}
\author{T. Verechtchaguina, I.M. Sokolov, and L. Schimansky-Geier}
\affiliation{Institute for Physics, Humboldt-University at Berlin,
Newton Str. 15, D-12489 Berlin, Germany}

\date{\today}

\begin{abstract}
Motivated by the dynamics of resonant neurons we discuss the properties of
the first passage time (FPT) densities for nonmarkovian differentiable
random processes. We start from an exact expression for the FPT density in terms
of an infinite series of integrals over joint densities of level crossings,
and consider different approximations based on truncation or on approximate
summation of this series. Thus, the first few terms of the series give good
approximations for the FPT density on short times. For rapidly decaying correlations
the decoupling approximations perform well in the whole time domain.

As an example we consider resonate-and-fire neurons representing stochastic
underdamped or moderately damped harmonic oscillators driven by white
Gaussian or by Ornstein-Uhlenbeck noise. We show, that approximations
reproduce all qualitatively different structures of the FPT densities: from
monomodal to multimodal densities with decaying peaks. The
approximations work for the systems of whatever dimension and are
especially effective for the processes with narrow spectral density,
exactly when markovian approximations fail.
\end{abstract}
\pacs{05.40.-a, 02.50.Ey, 87.17.Nn}
\maketitle

\setlength{\topmargin}{-1.3cm} \setlength{\textheight}{23.5cm} \draft

\section{Introduction}
The first passage time (FPT) is the time $T$ when a stochastic process $x(t)$
leaves an \textit{a priory} prescribed domain $\Delta$ of its state space
for the first time, assumed that $x(t)$ has been started at 
$t=0$ from a given initial value within $\Delta$. This concept was originally
introduced by E. Schr\"{o}dinger when discussing behavior of Brownian
particles in external fields \cite{S1915}. A large variety of problems ranging
from noise in vacuum tubes, chemical reactions and nucleation \cite{HTB1990}
to stochastic resonance \cite{LBM1991}, behavior of neurons \cite{T1988},
and risk management in finance \cite{R2001} can be reduced to FPT problems.
In majority of applications the attractor of the system's
dynamics lies inside $\Delta$. The escape process is characterized by the
noise-induced flux through the absorbing boundary of $\Delta$, i.e. by
the probability density $\mathcal{F}(T)$ of the first passage time.

Approaches to find $\mathcal{F}(T)$ are typically
based either on the Fokker-Planck equation with an absorbing
boundary \cite{R1989} or on the renewal theory \cite{vK1992}. Despite the
long history, explicit expressions for the FPT density are known only
for a few cases. These include overdamped particles under the influence of
white noise in the force free case, under time-independent constant forces
and linear forces  \cite{T1988,TM1977,GM1964,S1967}, as well as
the case of a constant force under colored noise \cite{B2004}.
Reasonable approximations exist for a few nonlinear forces
\cite{SH1989,LFS+}. The FPT densities of stationary markovian
processes have a very habitual form: $\mathcal{F}(T)$ goes through a single
maximum and then it decays either exponentially
or as a power law. Many neuronal systems do demonstrate such kind of
behavior.  It was shown already in \cite{GM1964}, that the interspike interval (ISI)
histograms obtained experimentally from output of some neurons can be
reproduced by FPT densities of one dimensional diffusion process.

This kind of description is suitable for overdamped systems, where
the relaxation time to the attractor $t_{rel}$ is much smaller than
the typical first passage time. At first the local
\textit{quasiequilibrium} is established in the system.
The escape occurs then from this equilibrium state and
follows with a constant rate $\kappa$ inversely proportional
to the mean FPT. This situation is closely related to the
Kramer's problem considering the quasistationary
flux over transparent boundary in the low noise limit.
The problem is independent of the detailed initial state and of
the time the trajectory has spent inside $\Delta$.
At times $T$ exceeding $t_{rel}$ the FPT probability density
decays exponentially: $\mathcal{F}(T) \sim \exp(-\kappa T)$. Well known
examples are chemical systems and nucleation processes,
where the rates determine the mean velocity of chemical
reactions or of forming overcritical nuclei \cite{HTB1990}.
Another example are the leaky integrate-and-fire and similar
neuronal models, where after the reset the corresponding
trajectories approach quickly the stable rest
state \cite{LGN+2004}.

However, if the time scale separation between the relaxation and
escape does not hold, the escape can occur before the establishment of the
quasiequilibrium and the rates are time dependent. The first passage time
depends sensitively on the initial conditions and the FPT densities have
a complex shape different from an exponential decay. In particular
this is the case on short time scales $T<t_{rel}$,
which attracted growing interest because of recent experiments studying chemical
reactions on time scales down to femtoseconds. The flux over the boundary
before the establishment of the quasiequilibrium was found to grow in a stepwise
manner, for an underdamped potential system staying initially at the bottom
of the well \cite{SSL+2001}. 

Our work is mainly motivated by dynamics of resonant neurons
\cite{MCD+1970,EKH+2004,BHR2003}. The voltage variable of
such a neuron exhibits damped subthreshold oscillations around
the attractive rest state. The characteristic relaxation time to the rest state is
large compared to the mean ISI. The escape of the voltage over the excitation
threshold is the beginning of a new spike. After spiking the voltage variable
is reset to a fixed value far from the rest state and then it can reach the
threshold \textit{prior} relaxation to the rest. The interspike interval
densities obtained from the output of resonant neurons show a
sequence of decaying peaks separated by intervals whose length is
of the order of the period of subthreshold oscillations.  In contrast to the Kramer's
rate theory we stress again the nonstationary character of this problem
due to the reset to sharp initial conditions. 

The multimodal ISI probability densities can be reproduced in models
with different mechanisms of subthreshold resonance: in Hodgkin-Huxley
model \cite{CWW2001}, in excitable FitzHugh-Nagumo model with the
stable fix point being a focus \cite{SL2002,VSS2004} or in the region
of a canard bifurcation \cite{VUZ+2003,MNV2001}. It was also shown that
two-component approximations might well mimic the spiking activity of
the stochastic Hodgkin-Huxley system \cite{T2005,TW2005}.
All these models have in common that the multimodal FPT density is
obtained for stochastic dynamical systems, which have at least two
dynamical variables, exhibit weakly, moderately damped or
self-amplifying oscillations and after a spike reset to initial values
which are not a fixed point. In the noise-free situation these systems
were denoted resonate-and-fire neurons \cite{I2001}. 

In the present work we aim to model excitable behavior with damped
subthreshold oscillations. First we present the general exact expression for
the FPT density for stochastic processes with differentiable trajectories
\cite{S1967,R1945,S1951,F1980}. It results in an infinite series of
integrals over joint densities of multiple level crossings. The later sum
stands for a sequential summing of trajectories excluding all except
the ones yielding the first passage. Furthermore, we discuss approximations
for FPT densities, which are based either on truncation of this series, or
on its approximate summation based on decoupling.
We prove the quality of different approximations by explicit calculations
for an underdamped harmonic oscillator driven by white or colored Gaussian
noise, representing the stochastic resonate-and-fire neurons.


\section{Exact expression for the first passage time density}
Consider a single random variable $x(t)$, whose $t$-dependence
is assumed to be differentiable. The first passage problem for $x(t)$
to a boundary $x_b$ is a special case of a level crossing problem.

The general theory of level crossings by a random process was
originally put forward by S.O. Rice \cite{R1945}. He derived an
expression for the probability density of recurrence of a stationary
random process to a given level in form of the so called Wiener-Rice
series \cite{S1951}. The exact expression for the first passage time
probability density Eq.(\ref {E:FPTmostGeneral}) is analogous to the
Wiener-Rice series and was discussed in \cite{F1980}, where the main result, our
Eq.(\ref{E:FPTmostGeneral}), was proved. We proceed by giving a much more
elementary derivation of Eq.(\ref{E:FPTmostGeneral}) which
serves as the main instrument in our further investigations.

Let us first discuss the probability $n_1(x_b,t|x_0,v_0)dt$ that a continuous
differentiable process $x(t)$ crosses the level $x_b$ in a time interval
between $t$ and $t+dt$ with positive velocity $v(t)=\dot{x}(t)>0$ under
initial conditions $x(0)=x_0,\dot{x}(0)=v_0$. Generally the whole
set of variables resulting from the markovian embedding of $x(t)$ should be
given at $t=0$. For simplicity we consider two dimensional dynamics, generalization
for the higher dimensional systems is obvious. Crossing the level
with positive velocity will be referred to as an upcrossing in what follows.

If $x(t)$ crosses the barrier within time interval $(t,t+dt)$ with velocity
$v>0$, then the value of coordinate at time $t$ should lie in interval
$x_b-vdt<x(t)<x_b$. The probability that $x(t)$ is in this interval equals
$\int \limits_{x_b-vdt}^{x_b}P(x,v,t|x_0,v_0,0)dx=|v|P(x_b,v,t|x_0,v_0,0)dt$.
Now, the velocity value at the instant of crossing is positive but
otherwise arbitrary. Thus we obtain the probability density of an upcrossing
by integration over all positive $v$: 
\begin{equation}
n_1(x_b,t|x_0,v_0,0)=\int\limits_0^{\infty }vP(x_b,v,t|x_0,v_0,0)dv.
\label{E:Rice}
\end{equation}
Eq. (\ref{E:Rice}) can be simply generalized to give the expression for
the joint probability density of multiple upcrossings. The probability
$n_p(x_b,t_p;\dots ;x_b,t_1|x_0,v_0,0)dt_p\dots dt_1$,
that the process $x(t)$ crosses the level $x_b$ in each of $p$ time intervals
$(t_1,t_1+dt_1),\dots, (t_p,t_p+dt_p)$ is given by: 
\begin{equation}
\begin{split}
n_p(x_b,t_p;\dots ;x_b,t_1|x_0,v_0,0)=\int\limits_0^{\infty }dv_p\dots
\int\limits_0^{\infty }dv_1&  \\
v_p\dots v_1 P(x_b,v_p,t_p;\dots ;x_b,v_1,t_1|x_0,v_0,0).& 
\end{split}
\label{E:RiceJoint}
\end{equation}

\begin{figure*}[tbh]
\begin{center}
\epsfig{file=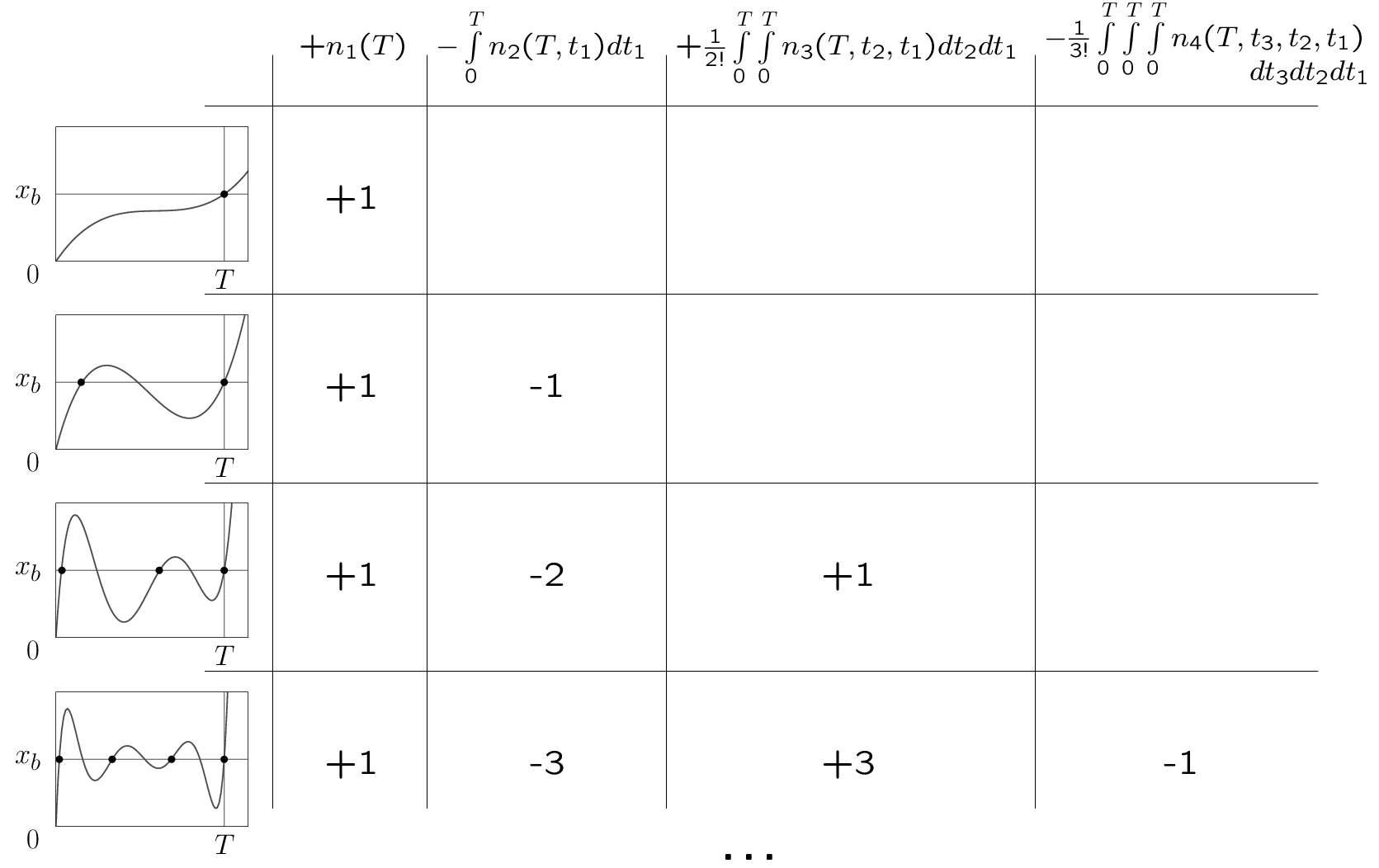,angle=90,width=13.5cm}
\end{center}
\caption{Counting crossings. The $N$th row corresponds to trajectories
with exactly $N$ upcrossings. The $p$th column corresponds
to the $p$th term of the sum Eq.(\ref{E:FPTgeneral}). The number on their
intersection gives, how many times a trajectory with exactly $N$ upcrossings
is accounted for in the $p$th term. Sum of all
numbers in every row is exactly zero.}
\label{F:CountingCrossings}
\end{figure*}
In what follows we omit $x_b$ and initial conditions in
expressions for the joint densities of upcrossings.
The transition probability densities are connected with joint
probability densities according to Bayes' theorem: 
\begin{equation}
\begin{split}
& P(x_p,v_p,t_p;\dots ;x_b,v_1,t_1| x_0,v_0,0)= \\
& \frac{P_{2p+2}(x_p,v_p,t_p;\dots ;x_b,v_1,t_;x_0,v_0,0)}{P_2(x_0,v_0,0)}.
\end{split}
\label{E:Bayes}
\end{equation}

Our aim now is to calculate the first passage time probability density
$\mathcal{F}(T)$, that is the fraction of all trajectories starting from the
initial point $x_0$ with initial velocity $v_0$ which perform the
upcrossing of the barrier at time $T$ and this upcrossing is the first one.
All such trajectories are accounted for in probability density $n_1(T)$
(see the first row in Fig. \ref{F:CountingCrossings}). However, $n_1(T)$
also accounts for trajectories for which the upcrossing at time $T$ was not the
first one, i.e. which had another upcrossing at some earlier time $t_1<T$
(row 2 in Fig. \ref{F:CountingCrossings}). Such trajectories should not
contribute to $\mathcal{F}(T)$, therefore we should substract them from
$n_1(T)$. Taking into account that $t_1$ can be arbitrary between $0$
and $T$, we get: 
\begin{equation}
n_1(T)-\int\limits_0^Tn_2(T,t_1)dt_1.
\label{E:2terms}
\end{equation}
This excludes all trajectories which cross $x_b$ exactly twice until $T$.
However, Eq.(\ref{E:2terms}) does not fully solve the problem since the
trajectories crossing $x_b$ three times, i.e at time $T$ and at two
earlier moments $t_i<T,i=1,2$ (row 3 in Fig. \ref{F:CountingCrossings}),
are not accounted for correctly. Each such trajectory is counted once in
$n_1(T)$. The second term in Eq.(\ref{E:2terms}) accounts for the pairs of
upcrossings at $T$ and at some $t_i<T$. Each trajectory with two additional
upcrossings at $t_i<T,i=1,2$ is therefore subtracted twice in
$\int_0^Tn_2(T,t_1)dt_1$. Such trajectories should not
contribute to $\mathcal{F}(T)$: in Eq. (\ref{E:2terms}) we subtracted too
much and have to add the amount of trajectories with three upcrossings at
times $0<t_i<T,i=1,2$ and $T$ again: 
\begin{equation}
n_1(T) -\int\limits_0^Tn_2(T,t_1)dt_1
+\frac{1}{2!}\int\limits_0^T\int \limits_0^Tn_3(T,t_2,t_1)dt_1dt_2.
\label{E:3terms}
\end{equation}
The factor $1/2!$ in the last term accounts for the number of permutations
of variables $t_i$.
Generally, if a trajectory crosses the level at time $T$ and at $N$ earlier
times $t_i<T,i=1,\dots ,N$, then in $\frac{1}{p!}\int \dots \int
n_{p+1}(T,t_p,\dots ,t_1)dt_1\dots dt_p$ it is accounted for exactly $C_N^p$
times ($C_N^p$ stands for the number of combinations). Note, that
$\sum \limits_{p=0}^{N}(-1)^{p}C_{N}^{p}=(1-1)^{N}=0$. Thus in the alternating
sum of the kind Eqs.(\ref{E:2terms}), (\ref{E:3terms}) containing $N+1$ terms,
all trajectories crossing $x_b$ at time $T$ and having  $i=1,2,\dots , N$ additional
upcrossings are excluded, however the ones with the larger
number of upcrossings are not accounted for correctly. Extending the sum to
infinity we exclude all superfluous trajectories, and only trajectories, for
which the upcrossing at time $T$ was the first one, remain. Thus the
expression for the first passage time probability density reads: 
\begin{equation}
\mathcal{F}(T)=\sum\limits_{p=0}^{\infty }\frac{(-1)^p}{p!}
\int\limits_0^T\dots \int\limits_0^T
 n_{p+1}(T,t_p,\dots,t_1)dt_p\dots dt_1 .
\label{E:FPTgeneral}
\end{equation}

Explicitly expressing the joint densities of upcrossings
using Eqs.(\ref{E:RiceJoint}),(\ref{E:Bayes}), we get
\begin{equation} \label{E:FPTmostGeneral}
\begin{split}
 \mathcal{F}&(T)  = \frac{1}{P_2(x_0,v_0,0)} \sum \limits_{p=0}^{\infty}
\frac{(-1)^p}{p!}
        \int \limits_0^T \dots \int \limits_0^T  dt_p \dots dt_1 \\
     &   \int \limits_0^{\infty} 
        \dots  \int \limits_0^{\infty}dv dv_1 \dots dv_p v v_1 \dots v_p \\
     &  P_{2p+4}
        (x_b,v, T;x_b,v_p,t_p; \dots; x_b,v_1, t_1;x_0,v_0,0).
\end{split}
\end{equation}

Eq.(\ref{E:FPTgeneral}) connects $\mathcal{F}(T)$, i.e. the solution of the FPT
problem with absorbing boundary at $x_b$, with all joint densities of upcrossings for
the unbounded process. To obtain $n_p(t_p, \dots, t_1)$  we consider trajectories,
which are not absorbed at $x_b$, but can return after an upcrossing and
then cross $x_b$ again and again. The right combination of all these densities
of multiple level crossings results in the probability density for the first upcrossing.

An alternative to direct summation of infinite series Eq.(\ref{E:FPTgeneral})
is based on an analogue to a cumulant expansion.
The times, when the random process $x(t)$ performs upcrossings of $x_b$
form a \textit{point process}, or a \textit{system of random points}
\cite{S1967,vK1992}. The functions
\begin{equation} \label{E:PointProcDistr}
n_1(t_1), \qquad n_2(t_2,t_1), \qquad n_3(t_3,t_2,t_1), \qquad \dots 
\end{equation}
are the \textit{distribution functions} of the point process.
Since $x(t)$ has finite velocity, the interval between two upcrossings
can not be arbitrary small and so
$n_p(t_p,\dots,t_1)$ vanishes if two of its arguments coincide. Such
random point process is called a \textit{system of nonapproaching points} \cite{S1967}.
In this context $\mathcal{F}(T)$ is interpreted as the waiting-time density
of the point process. The last is the probability density for the time $T$
when the first event occur.

The system of random points is completely characterized by its cumulant functions
\begin{equation} \label{E:PointProcCorrF}
g_1(t_1), \qquad g_2(t_2,t_1),  \qquad g_3(t_3,t_2,t_1), \qquad \dots .
\end{equation}
Choose an arbitrary natural number $r$, and then fix $r$ arbitrary
numbers $z_1,\dots, z_r$ and $r$ positive times $t_1, \dots , t_r$. The
cumulant functions are then defined by the relation
\begin{equation} \label{E:CorrFuncGeneral}
\begin{split}
 1+\sum_{p=1}^{\infty} \frac{1}{p!} \sum_{\alpha, \dots, \omega=1}^r
n_p(t_{\alpha},\dots,t_{\omega})z_{\alpha}\dots z_{\omega} = \\
\exp \left( \sum_{p=1}^{\infty} \frac{1}{p!} \sum_{\alpha, \dots, \omega=1}^r
g_p(t_{\alpha}, \dots , t_{\omega})z_{\alpha} \dots z_{\omega} \right).
\end{split}
\end{equation}

We obtain the explicit relations between cumulant and distribution functions
of the point process if we differentiate both sides of Eq.(\ref{E:CorrFuncGeneral})
over all $z_i$ and then set $z_i=0,(i=1, \dots, r)$, i.e. if we apply the operator
$\partial^r/(\partial z_1 \dots \partial z_r) \big|_{z_1=_{\dots}=z_r=0}$.
Doing so sequentially for $r=1,2,3,\dots$ we get
\begin{equation}
\begin{split}
& g_1(t_1) =n_1(t_1), \\
& g_2(t_2,t_1) =n_2(t_2,t_1)-n_1(t_1)n_1(t_2), \\
& g_3(t_3,t_2,t_1) = n_3(t_3,t_2,t_1)-3\{ n_1(t_1)n_2(t_3,t_2)\}_s \\
& \qquad \qquad \qquad +2n_1(t_1)n_1(t_2)n_1(t_3), \qquad \dots
\end{split}
\label{E:CorrFuncSpecial}
\end{equation}
Here $\{ \dots \}_s$ denotes the operation of symmetrization of the expression
in the brackets with respect to all permutations of its arguments. The coefficients
in these forms are the same as in relations between the moments and the
cumulants of  a random variable.

The relation Eq.(\ref{E:CorrFuncGeneral}) does not change its form
if we choose different times $t_1,\dots , t_r$, different values
$z_1, \dots , z_r$ or change the number $r$.
Thus extending $r$ to infinity, allowing $t$ to take all possible
values between $0$ and $T$, and choosing $z_1=_{\dots}=z_r=-1$
we get from Eq.(\ref{E:CorrFuncGeneral})
\begin{eqnarray} \label{E:CorrFuncIntegral}
&1 +{\displaystyle \sum \limits_{p=1}^{\infty} \frac{(-1)^p}{p!} \int \limits_0^T \dots \int \limits_0^T
n_p(t_p,\dots,t_1) dt_p \dots dt_1 =} & \nonumber \\
\\
 & \exp \left( {\displaystyle \sum \limits_{p=1}^{\infty} \frac{(-1)^p}{p!}  \int \limits_0^T \dots \int \limits_0^T
g_p(t_p, \dots , t_1)  dt_p \dots dt_1 } \right). & \nonumber
\end{eqnarray}

It is easy to verify, that the derivative $(-d/dT)$ of the expression
on the left hand side of Eq.(\ref{E:CorrFuncIntegral}) is exactly
the expression on the right hand side of Eq.(\ref{E:FPTgeneral}).
Thus differentiating the right hand side of
Eq.(\ref{E:CorrFuncIntegral}) over $T$ we obtain the expression for
the waiting-time density $\mathcal{F}(T)$ through the cumulant
functions of the point process:
\begin{equation}
\mathcal{F}(T) = S^{\prime}(T)e^{-S(T)} \label{E:GenForm}
\end{equation}
with 
\begin{equation}
S(T)= - \sum_{p=1}^{\infty }\frac{(-1)^p}{p!}
\int \limits_0^T\dots \int \limits_0^Tg_p(t_p,\dots ,
t_1) dt_p \dots dt_1.\label{E:Sexact}
\end{equation}

Eqs.(\ref{E:PointProcCorrF})-(\ref{E:CorrFuncIntegral}) are general
expressions which hold for systems of random points, defined by
distribution functions Eq.(\ref{E:PointProcDistr}) of any kind.
So are also Eqs.(\ref{E:FPTgeneral}) and (\ref{E:GenForm}), (\ref{E:Sexact}),
which give the waiting-time density for arbitrary point process.
In particular, for the random points being the times, when a
differentiable random process crosses the level $x_b$,
the distribution functions $n_p(t_p,\dots ,t_1)$
are given by the joint densities of upcrossings Eq.(\ref{E:RiceJoint}).
Then Eqs.(\ref{E:FPTgeneral}) and (\ref{E:GenForm}), (\ref{E:Sexact})
together with Eq.(\ref{E:RiceJoint}) express the first passage time density
for this differentiable random process. The function $S^{\prime }(T)$ can
be interpreted as a the time-dependent escape rate.

These are the exact results for the FPT probability density of any continuous
differentiable random process. Though these results were employed for
mathematical proofs, up to our knowledge these infinite series of multiple
integrals was never
used for explicit calculations. We proceed to show, that Eqs.(\ref{E:FPTgeneral})
and (\ref{E:GenForm}), (\ref{E:Sexact}) can be a starting point for several
approximations. As often in the case of infinite series, the useful approximations
can be based either on the truncation of the series after several
first terms calculated exactly, or by approximation of the higher
order terms through the lower order ones what might lead to a closed analytical
form. Truncation approximations for Eq.(\ref{E:FPTgeneral}) are not
normalized, hold only on short time scales, and diverge at longer
times (due to the miscount of trajectories with several upcrossings).
The approximations of the second type are based on a subsummation
in Eq.(\ref{E:Sexact}) for $S(T)$. They are normalized and can be
used in the whole time domain.  Note, only approximations guaranteeing
positive rates $S^{\prime}(T)$ are reasonable. Thus the set of possible
approximations for the series Eq.(\ref{E:Sexact}) is rather restricted.


\section{Noisy driven harmonic oscillator: resonate-and-fire}\label{S:model}
The model we have
in mind is the resonate-and-fire model of a neuron \cite{I2001}. This is
the least complicated model accounting for the resonant properties
of neurons in terms of an equivalent RLC circuit \cite{MCD+1970, EKH+2004}.
In this way it is directly related to the leaky integrate-and-fire model also
based on the electrical analogy. Alternatively the model can be
interpreted as a systematic and linearized reduction of Hodgkin-Huxley
type dynamics \cite{K1999}. For the sake of simplicity we neglect
the absolute refractory time. Under this assumption the model is equivalent to
the underdamped harmonic oscillator with the threshold and reset,
what makes the results applicable in many other domains of science.
We change by time scale and variable transformations to dimensionless
parameters and variables. The dynamics of the voltage variable $x(t)$ is given by:
\begin{equation} \label{E:HarmOscil} 
\dot{x} =v;  \qquad \dot{v} = -\gamma v-\omega _{0}^{2}x+\eta (t).
\end{equation}
We fix the frequency $\omega _0=1$,  choose initial conditions for $x$ and
its velocity $v$ to be $x_0=-1,v_0=0$ and set the threshold at $x_b=1$.
In the present paper we consider two types of noisy drive:
(i) the white noise $\eta (t)=\sqrt{2D}\xi (t)$, and (ii)
the Ornstein-Uhlenbeck noise $\dot{\eta}=-\tau ^{-1}\eta +\sqrt{2D}\tau
^{-1}\xi (t)$, with $\xi (t)$ being the white Gaussian noise of intensity 1.
First we concentrate on the case (i) of white noise driving.

Because of the linearity of the system Eq.(\ref{E:HarmOscil}) all joint
probability densities are Gaussian and have the form \cite{R1989}:
\begin{equation}
P_{n}(\vec{Q})=\frac{1}{(2\pi )^{n/2}\sqrt{\det \hat{C}_{n}}}\exp \left( -
\frac{\vec{Q}\hat{C}_{n}^{-1}\vec{Q}}{2}\right) .  \label{E:GaussProbab}
\end{equation}
Here $\vec{Q}=(q_{1}(t_{1}),\dots ,q_{n}(t_{n}))$ is an $n$-dimensional
vector, whose $i$th component is the value of coordinate $x(t_i)$ or of
velocity $v(t_i)$ at the moment $t_i$. $\hat{C}_{n}$ is symmetric
$n\times n$ correlation matrix. Its elements are correlation functions
between corresponding components of vector $\vec{Q}$ : $
c_{ij}=c_{ji}=\langle q_{i}(t_{i})q_{j}(t_{j}) \rangle$.

Correlation functions for the system Eq.(\ref{E:HarmOscil}) are easily
obtained using Fourier transform and Wiener-Khinchin theorem \cite{R1989}.
For the case of white noise driving and in an underdamped regime ($\gamma<2\omega_0$)
$r_{xx}(t)=\langle x(t^{\prime})x(t^{\prime}+t) \rangle =
\frac{D}{\gamma \omega_0^2}e^{-\frac{\gamma}{2}t}
\left( \frac{\gamma}{2\Omega}\sin(\Omega t)+\cos(\Omega t) \right)$,
with $\Omega=\sqrt{|\omega_0^2-\frac{\gamma^2}{4}|}$.
In overdamped case ($\gamma>2\omega_0$) the expression for $r_{xx}(t)$ is
the same, except the trigonometric functions are replaced with hyperbolic ones.
Further, $r_{xv}(t)=r'_{xx}(t)$ and $r_{vv}(t)=-r''_{xx}(t)$.

Then $n_1(T)$ is obtained from Eq.(\ref{E:Rice}) in closed analytical form: 
\begin{equation}
\begin{split}
n_1& (T)=\frac{\sigma _{x}\sigma _{v}}{2\pi \mu _{22}\sqrt{\det \hat{C}
_{4}}}\exp \left[ \frac{\sigma _{x}^{2}v_{0}^{2}+\sigma _{v}^{2}x_{0}^{2}}{
2\sigma _{x}\sigma _{v}}\right]  \\
& \exp \left[ -\frac{1}{2}\sum\limits_{i,j\neq 2}\mu _{ij}q_{i}q_{j}\right]
\left[ 1-\sqrt{\pi }\alpha e^{\alpha ^{2}}\mathrm{erfc}(\alpha )\right] .
\end{split}
\label{E:1termExpl}
\end{equation}
Here $\mu _{ij}=\mu _{ji}$ are elements of the inverse correlation matrix
$(\hat{C}_{4})^{-1}$, $q_{i}$ are components of the vector $\vec{Q}
=(x(T),v(T),x_{0}(0),v_{0}(0))$. The dispersions of $x$ and
$v$ are $\sigma_x^2=r_{xx}(0)=D/\gamma \omega_0^2$,
$\sigma_v^2=r_{vv}(0)=D/\gamma$. We have introduced
$\alpha =\left(\sum\limits_{i\not{=}2}\mu _{2i}q_{i}\right) /\sqrt{2\mu _{22}}$.
Finally, $\mathrm{erfc}(x)$ is a complementary error function.

For the joint densities
of multiple upcrossings $n_p(t_p, \dots , t_1)$ no closed expressions can
be obtained. We evaluate the integral over $v_1$ in Eq.(\ref{E:RiceJoint})
analytically and  then perform numerical integration of the resulting expression
over $v_2, \dots, v_p$ to obtain $n_p(t_p, \dots , t_1)$. The integrals over time
in the expressions for $\mathcal{F}(T)$ are also evaluated numerically.


\section{Truncation approximations}\label{S:Trunc}

\begin{figure}[tbh]
\begin{center}
\epsfig{file=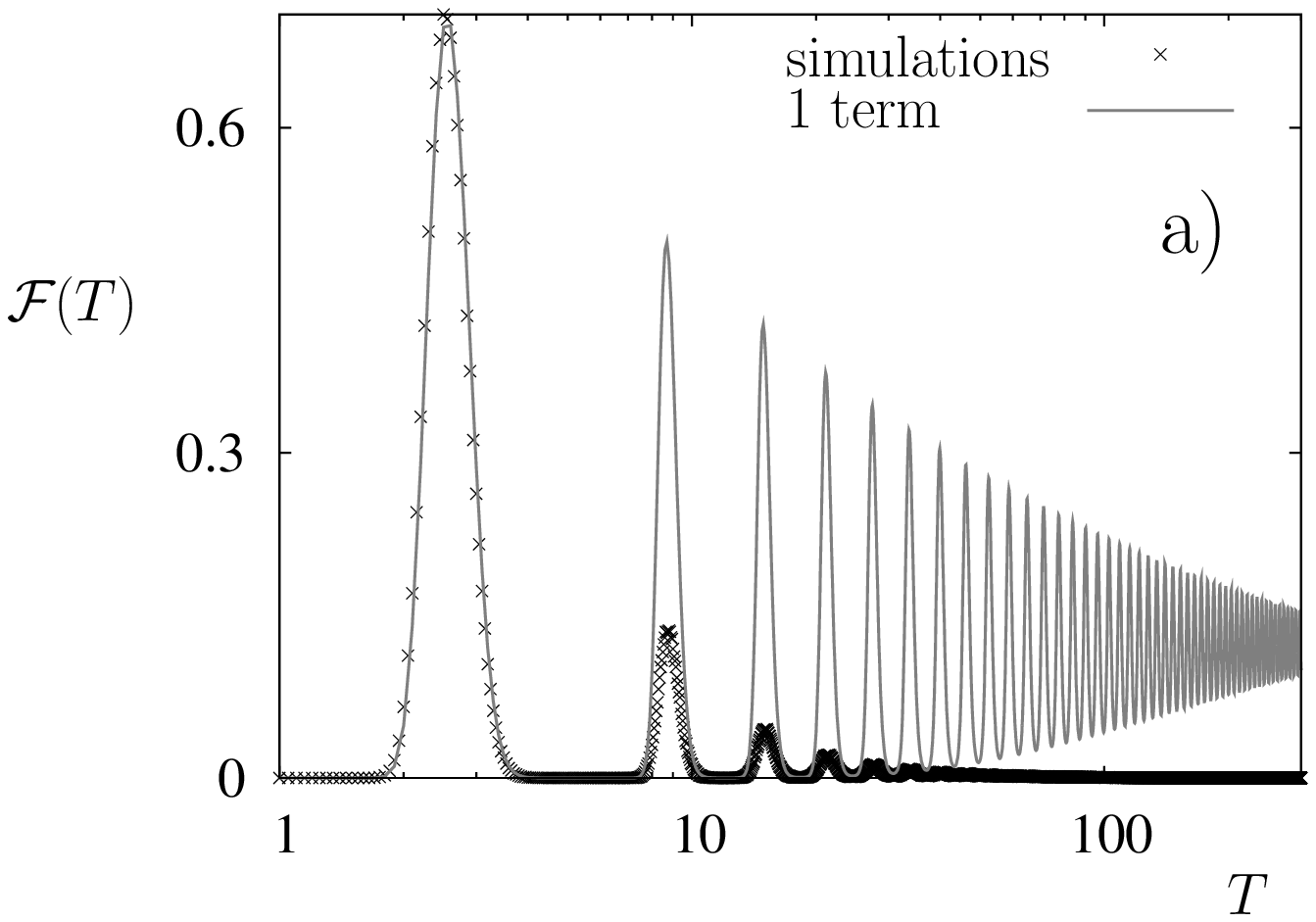,width=7.5cm} 
\epsfig{file=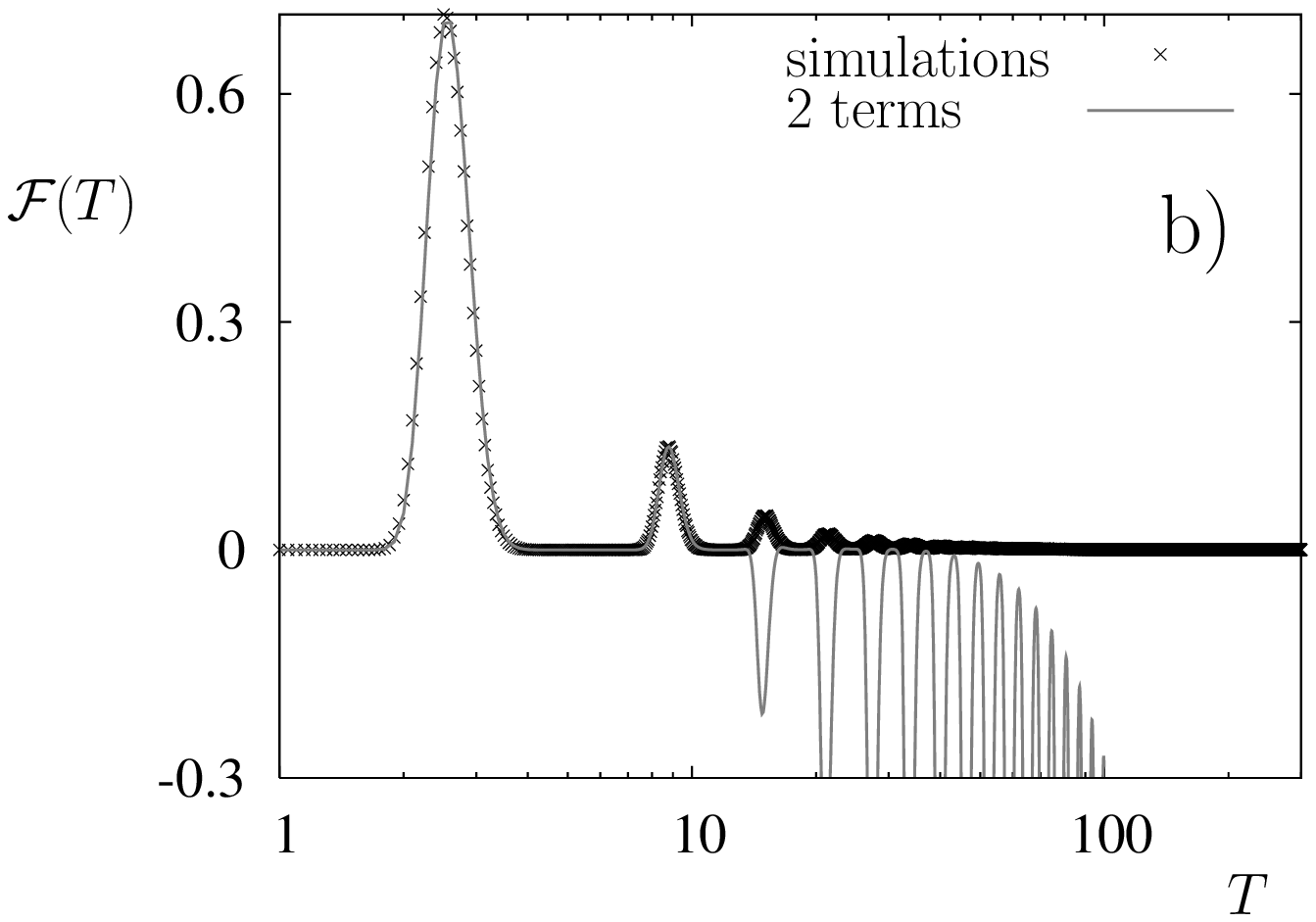,width=7.5cm}
\epsfig{file=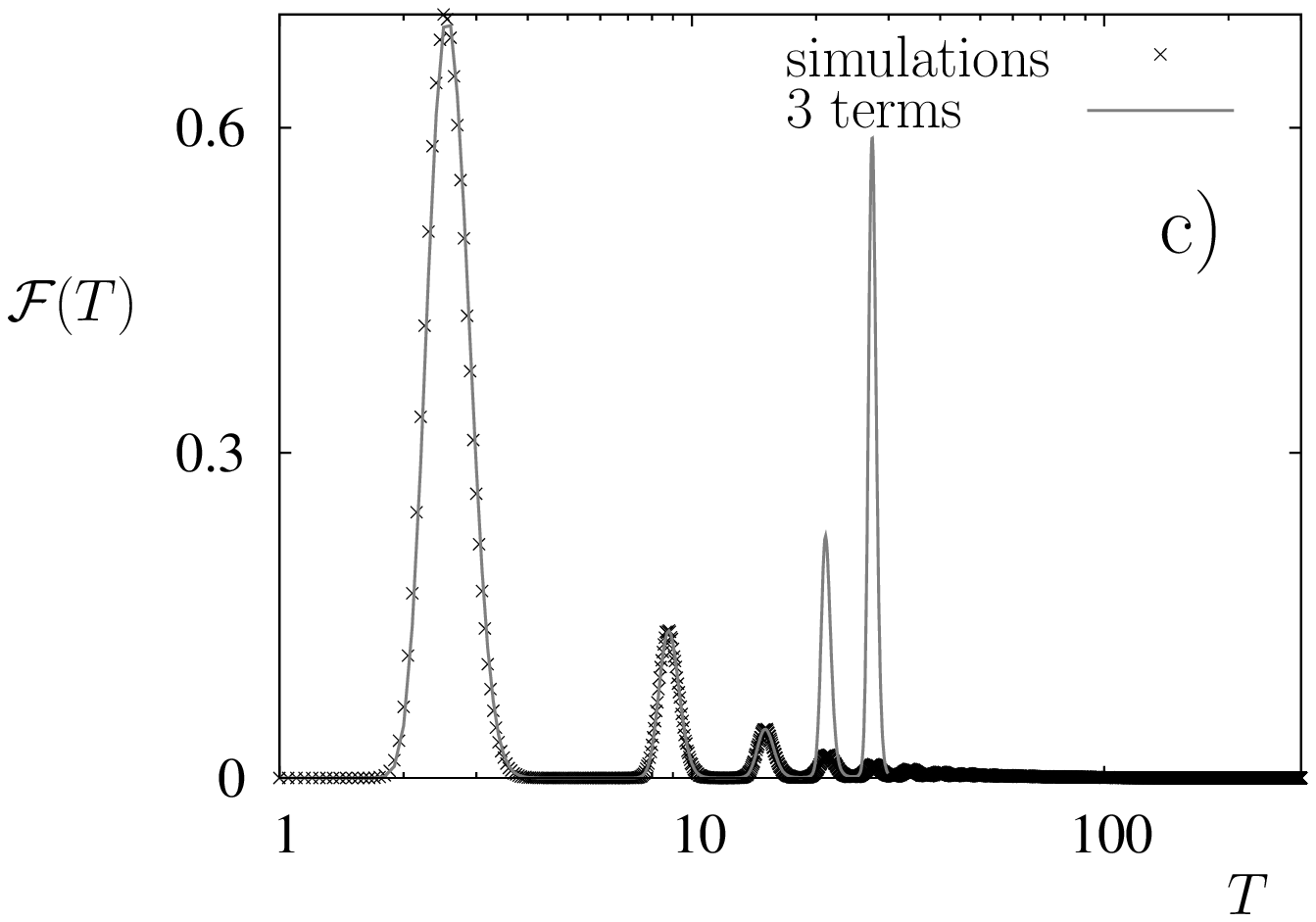,width=7.5cm}
\end{center}
\caption{FPT density for harmonic oscillator driven by white Gaussian noise,
$\omega_0=1, \gamma=0.01, D=0.02, x_0=-1, v_0=0, x_b=1$. Simulation
results shown with black crosses and truncation approximations with gray
line: (a) 1 term Eq.(\ref{E:Rice}), (b) 2 terms Eq.(\ref{E:2terms}) and
(c) 3 terms Eq.(\ref{E:3terms}). The mean FPT obtained from simulations
equals $14.6$, the median of distribution lies by $3.2$. Note the
logarithmic scale.}
\label{F:FPTG0.01D0.02}
\end{figure}

The first passage time density $\mathcal{F}(T)$ for the harmonic oscillator
driven by white Gaussian noise obtained from simulations is depicted with
black crosses in Fig.\ref{F:FPTG0.01D0.02}. Parameters are chosen to be
$\gamma =0.01,D=0.02$. In this case of very
small friction the correlation functions of the process oscillate
with period $T_{p}=2\pi /{\sqrt{\omega_0^2-\gamma^2/4}}$ and decay
slowly within the relaxation time $t _{rel}=2/\gamma $.
A typical trajectory is smooth and shows almost regular
oscillations with fluctuating phase and amplitude. The probability to reach
$x_b$ is higher in the maxima of the subthreshold oscillations. The initial phase of
these oscillations is fixed by sharp initial conditions. Thus on shorter time
scales $\mathcal {F}(T)$ shows the multiple peaks following with the frequency
of damped oscillations $\Omega=\sqrt{\omega_0^2-\gamma^2/4}$. On long times
$T\gg t_{rel}$ the quasiequilibrium establishes and FPT density decays
exponentially. The number of visible peaks depends on the
relation between $t_{rel}$ and the period of oscillations $T_p$ and is given by
the number of periods elapsing before the quasiequilibrium is achieved.
For parameter values as in Fig.\ref{F:FPTG0.01D0.02}  $t_{rel}=200$, what
corresponds to about 30 periods $T_p=6.28$.

Let us consider truncation approximations for the series Eq.(\ref{E:FPTgeneral}).
The first approximation is given by the first term $n_1(T)$, the second approximation
by 2 terms, Eq.(\ref{E:2terms}), and the third by 3 terms,
Eq.(\ref{E:3terms}). The higher order approximations entail the numerical
estimation of highdimensional integrals, which at some stage leads to a
computational effort larger than the one necessary for a direct simulation.
Therefore we restrict ourselves to the 1-, 2- and 3- terms approximations.

The result of the 1 term approximation is shown in Fig.\ref{F:FPTG0.01D0.02}(a)
with a gray line. The first peak of the FPT density is
reproduced almost exactly. All further peaks are overestimated, because all
trajectories performing multiple upcrossings of $x_b$ are included.
On long times the process becomes stationary and the first approximation tends
to a constant value $\lim_{T \rightarrow \infty}n_1(T)=n_0$.  This is the
mean frequency of upcrossings for a stationary process, also known as the
Rice frequency \cite{R1945}. The general expression for $n_0$ reads
\begin{equation}
n_0=\frac{1}{2\pi }\left(- \frac{r''_{xx}(0)}{r_{xx}(0)}\right)
^{1/2}e^{-x_{b}^{2}/2r_{xx}(0)}.
\end{equation}
\label{E:FreqCrossHO}
In our case of a harmonic oscillator driven by white noise
$n_0=(\omega_0/2\pi)\exp{(-\gamma x_b^2 \omega_0^2/2D)}$.
In the stationary regime the mean interval between two consecutive
upcrossing $T_R$ is given by the inverse of the Rice frequency
$T_R=1/n_0$. For the chosen parameter values $T_R=8.06$.

The second approximation (gray line in Fig.\ref{F:FPTG0.01D0.02}(b))
reproduces almost exactly the first two peaks of FPT density. Then it
becomes negative, because in Eq.(\ref{E:2terms}) trajectories performing 2
and more superfluous upcrossings are subtracted too many times.
Moreover the second approximation tends to minus infinity for
$T\rightarrow \infty $. The third approximation reproduces well the three
first peaks of $\mathcal{F}(T)$, and then diverges tending to plus infinity.

Note that the mean first passage time
obtained numerically equals $14.6$ for these parameter values, and
the median of the distribution lies by $3.2$. Thus the first three approximations
reproduce the most part of the FPT probability density.

\begin{figure}[tbh]
\begin{center}
\epsfig{file=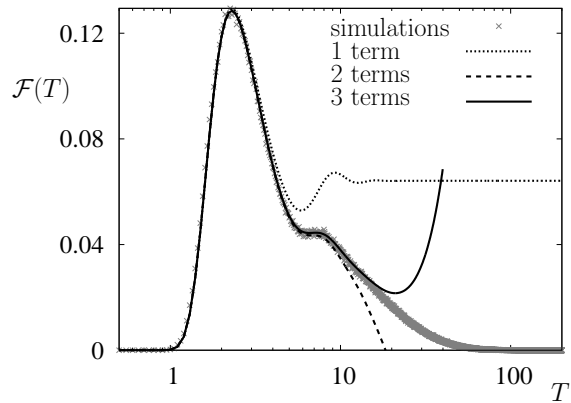,width=7.5cm}
\end{center}
\caption{FPT density for harmonic oscillator driven by white Gaussian noise,
$\omega_0=1, \gamma=0.8, D=0.44, x_0=-1, v_0=0, x_b=1$. Simulation results
shown with gray crosses; 1 term approximation Eq.(\ref{E:Rice}) with a
dot line, 2 terms approximation Eq.(\ref{E:2terms}) with a dashed line,
and 3 terms approximation Eq.(\ref{E:3terms}) with a solid line. The
mean FPT obtained from simulations equals $13.1$, the median of distribution
lies by $9.1$. Note the logarithmic scale.}
\label{F:FPTG0.80D0.44}
\end{figure}

The behavior of $\mathcal{F}(T)$ for the harmonic oscillator with higher
damping $\gamma=0.8$, stronger noise intensity $D=0.44$, and other
parameters as in Fig.\ref{F:FPTG0.01D0.02} is presented in
Fig.\ref{F:FPTG0.80D0.44}. For these parameter values the relaxation
time $t_{rel}=2.5$ is less than the period $T_p=6.86$. Therefore the FPT density is
practically monomodal with a single maximum and a small shoulder separating
it from the exponential tail. The numerically obtained $\mathcal{F}(T)$ is shown
with gray crosses, the 1 term truncation with a dot
line, the 2 terms truncation with a dashed line, and the 3
terms truncation with a solid line. The truncation
approximations reproduce again the most part of the distribution: the mean
FPT equals 13.1 and the median lies by 9.1.

\begin{figure}[tbh]
\begin{center}
\epsfig{file=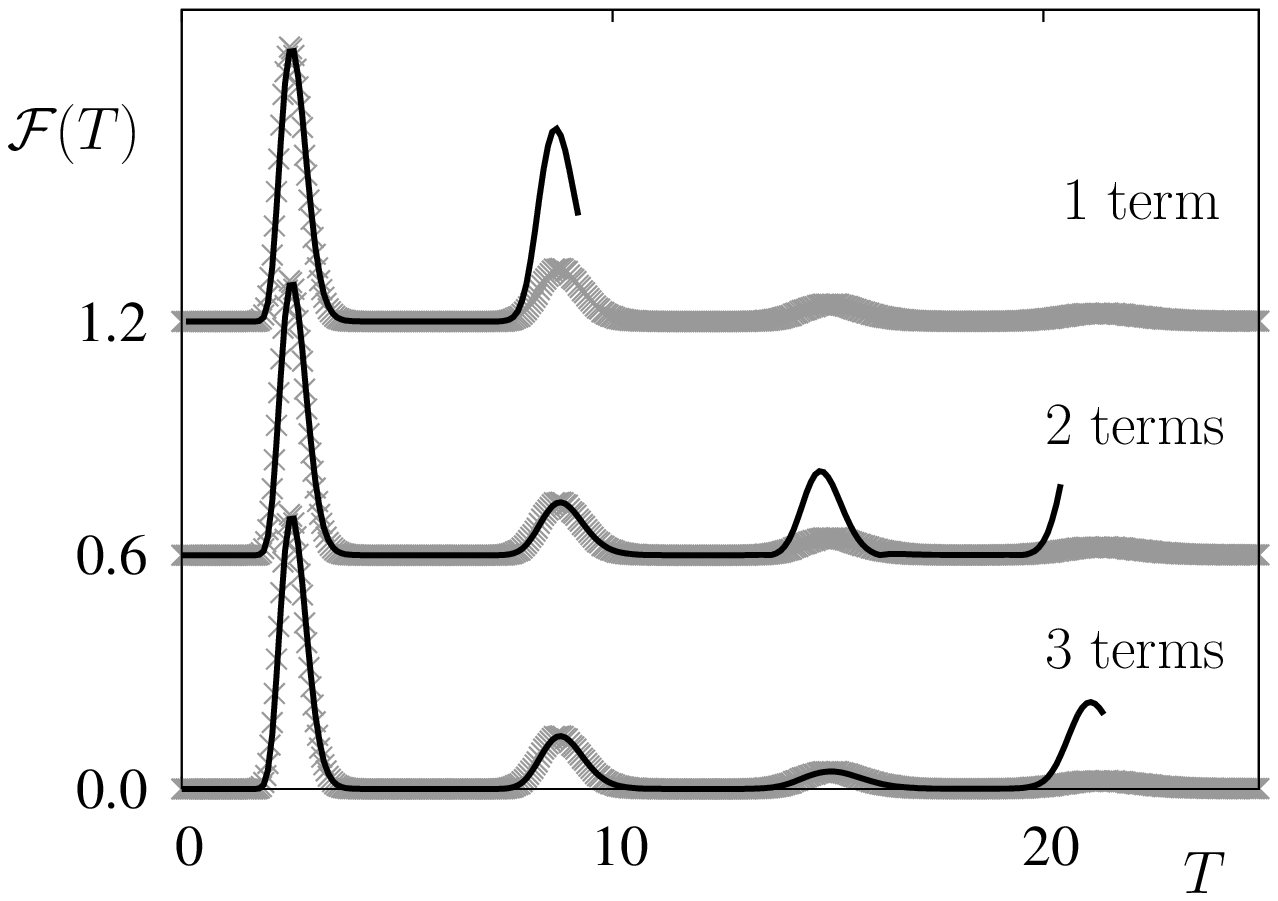,width=7cm}
\end{center}
\caption{The same as in Fig.\ref{F:FPTG0.01D0.02}, however the
truncations are correctly normalized by Eq.(\ref{E:TruncNorm}). Simulation results
shown with gray crosses; the truncations with solid lines.
The curves for normalized 1,2, and 3 terms truncations are vertically shifted
by 0.6 for the sake of clarity.}
\label{F:FPTG0.01D0.02LSG}
\end{figure}

Thus the truncation approximations reproduce the FPT density on short times but
are not normalized and diverge on large times. However, one can force the
normalization in truncations as follows
\begin{equation}\label{E:TruncNorm}
\left|\mathcal{F}_{trunc}(T)\right|\Theta \left(1-\int_0^T \left|\mathcal{F}_{trunc}(t)\right|dt \right).
\end{equation}
Here $\mathcal{F}_{trunc}(T)$ is a truncated series for the FPT
density, it is given by Eqs.(\ref{E:Rice}),(\ref{E:2terms}),
and (\ref{E:3terms}) for 1,2, and 3 terms truncations respectively.
$\Theta(t)$ is the Heaviside step function: the expression
Eq.(\ref{E:TruncNorm}) turns to zero as soon as the integral over the
absolute value of $\mathcal{F}_{trunc}(T)$ exceeds 1.

In Fig.\ref{F:FPTG0.01D0.02LSG} we show the FPT density
for the same parameter values as in Fig.\ref{F:FPTG0.01D0.02}.
Simulation results are shown with crosses, and approximations obtained
from Eq.(\ref{E:TruncNorm}) with
1,2, and 3 terms by solid lines. The probability distributed over the long
exponential tail in the real FPT density, is concentrated in the
positive artefact posed on intermediate times in the normalized truncations.
Hence the mean FPT computed from such approximations is always
strongly underestimated.
 
The more terms are included, the more precise truncations become.
However one has to confine oneself to a few terms, since the calculation of
higher order terms implies the computation of multiple integrals
and is not more effective than simulations.
Therefore the truncation approximations are good, when the most part of the
FPT probability is concentrated in the first few peaks, i.e. when the barrier
value is low or the noise is strong.


\section{Decoupling approximations}

Decoupling approximations for Eq.(\ref{E:FPTgeneral}) or Eq.(\ref{E:Sexact})
are based on approximate expressions of the
higher order terms through the lower order ones, what may lead to a closed
analytical form. Thereby infinitely many approximate terms are included.

The simplest way to obtain such an approximation is to neglect all correlations
between upcrossings. This means to neglect all terms in Eq.(\ref{E:Sexact})
except for the first one, and leads to
\begin{equation}
S(T)=\int \limits_0^T n_1(t)dt.  \label{E:Hertz}
\end{equation}

Equivalently, neglecting
all correlations corresponds to the factorization of $n_{p+1}(T,t_p,\dots,t_1)$ into
a product of one-point densities $n_1(T)n_1(t_p) \dots n_1(t_1)$ in
Eq.(\ref{E:FPTgeneral}). Then the series, Eq.(\ref{E:FPTgeneral}) sums up
into $\mathcal {F}(T)\approx n_1(T)\exp \left(-\int_0^T n_1(t)dt\right)$, which is
equivalent to Eqs.(\ref{E:GenForm}),(\ref{E:Hertz}).
This approximation will be refereed to as the \textit{Hertz approximation} since 
the form of ${\cal F}(T)$ resembles the Hertz distribution \cite{H1909}.
It is an approximation of first order, since it takes
the first term of the series exactly, and all other terms are approximated
through this first one.

\begin{figure}[tbp]
\begin{center}
\epsfig{file=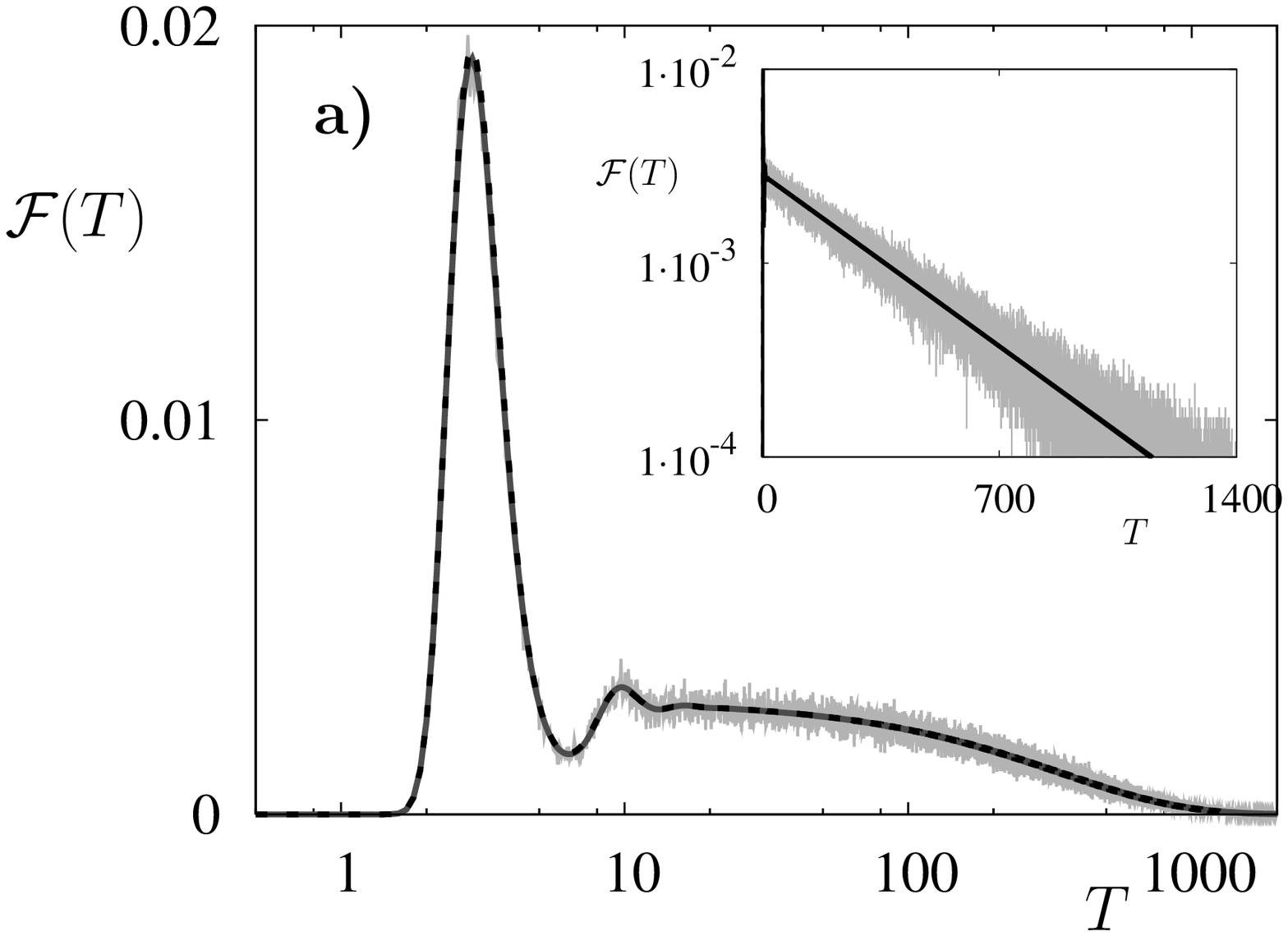,width=7.5cm}
\epsfig{file=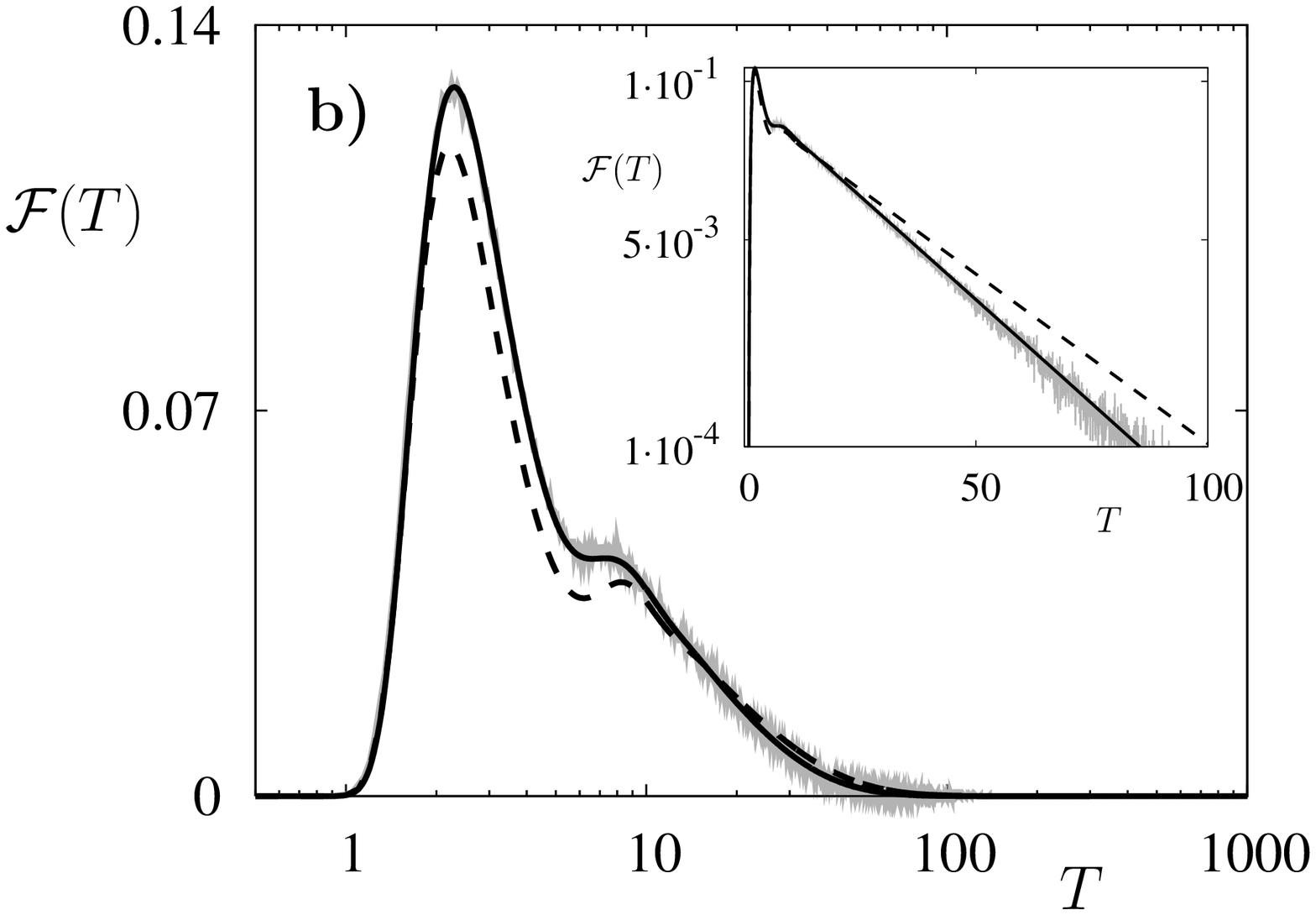,width=7.5cm}
\epsfig{file=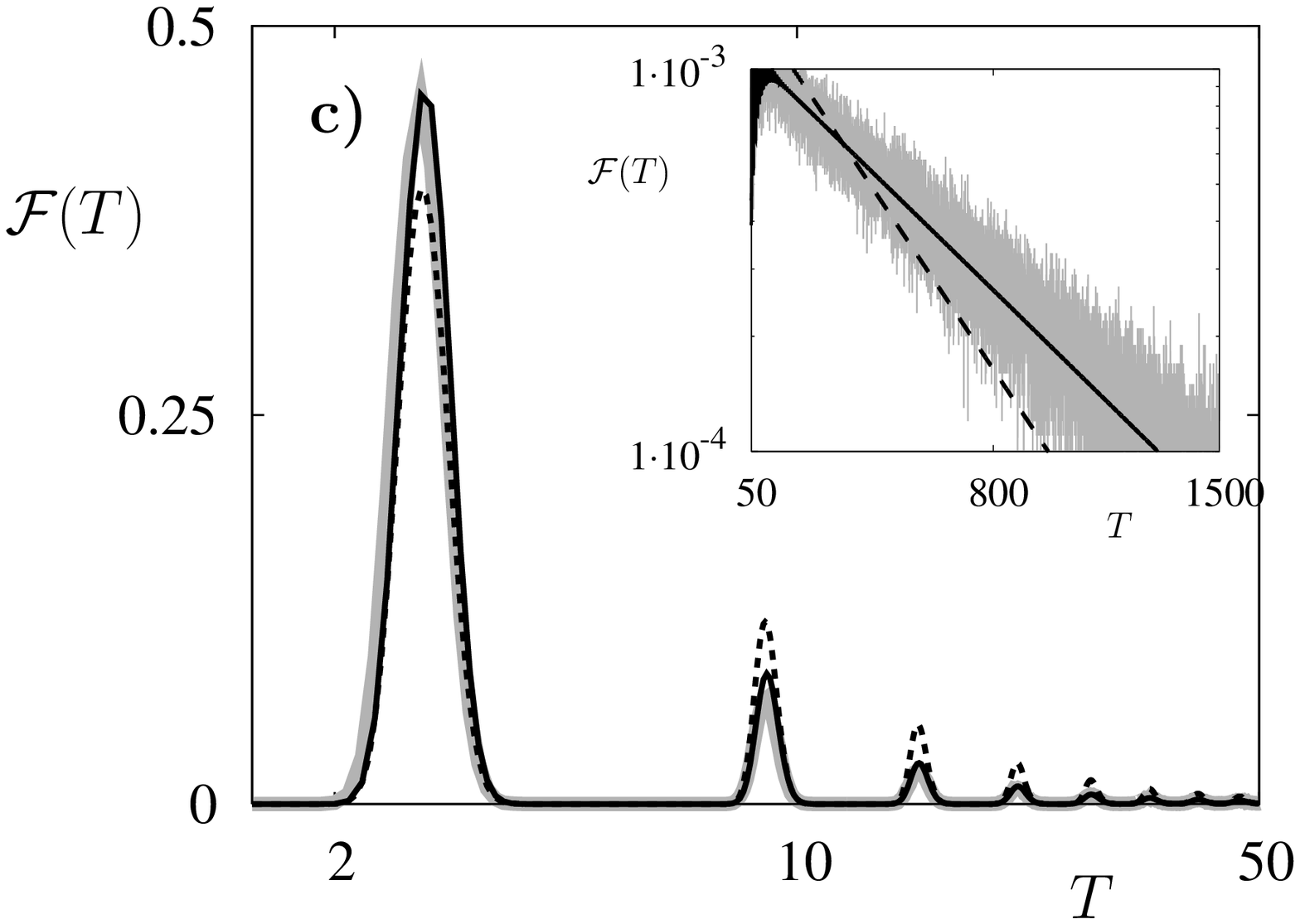,width=7.5cm}
\end{center}
\caption{FPT probability density for harmonic oscillator driven by Gaussian
white noise. Simulation results are shown with a gray line, Hertz
approximation with a black dashed line, and Stratonovich approximation with
a black solid line. Note the logarithmic scale in $T$. The insets show the
same curves on the logarithmic scale in ${\cal F}(T)$. The parameters
are: $\omega_0=1,x_0=-1,v_0=0,x_b=1$, a) $\gamma = 0.8, D=0.1,
t_{rel}=2.5, T_R=343$, b) $\gamma=0.8, D=0.44, t_{rel}=2.5, T_R=15.6$,
c) $\gamma=0.08, D=0.01, t_{rel}=25, T_R=343$}
\label{F:FPTunderdamped}
\end{figure}

The second order approximation should therefore account the first and
the second terms exactly and approximate all higher terms through these two.
The general form of $\mathcal{F}(T)$ in terms of the cumulant functions
Eq.(\ref{E:GenForm}) ensures the right normalization, irrespective
of the way how $S(T)$ is approximated. However the simple truncation of the
series Eq.(\ref{E:Sexact}) after the second term does not ensure
the positive escape rate $S^{\prime}(T)$.

The second order approximation guaranteeing $S^{\prime}(T)>0$
was proposed by Stratonovich in the context of peak duration \cite{S1967}.
The first and the second cumulant functions are taken exactly,
and the higher ones are approximated by the combinations of these two:
\begin{equation}
\begin{split}
g_p(t_p,\dots,t_1) \approx &  (-1)^{p-1}(p-1)! n_1(t_p) \dots n_1(t_1) \\
& \{ R(t_1,t_2)R(t_1,t_3) \dots R(t_1,t_p) \}_s.
\end{split}
\label{E:CorrFuncApp}
\end{equation}
Here $\{ \dots \}_s$ is again the operation of symmetrization. $R(t_i,t_j)$
is the correlation coefficient of upcrossings
\begin{equation}
R(t_{i},t_{j})=1-\frac{n_2(t_{i},t_{j})}{n_{1}(t_{i})n_{1}(t_{j})}.
\label{E:CorCoef}
\end{equation}
Note, that $R(t_1,t_1)=1$ and $R(t_i,t_j) \rightarrow 0$ for large
values of $|t_i-t_j|$.

The approximation of the cumulant functions in form Eq.(\ref{E:CorrFuncApp})
can be motivated by the following argument. Consider
Eq.(\ref{E:CorrFuncGeneral}) with $r=1$. Recall that the joint densities
of upcrossings vanish for coinciding arguments: $n_p(t_1, \dots , t_1)=0$.
Thus it follows from Eq.(\ref{E:CorrFuncGeneral}):
\begin{equation}
\ln(1+n_1(t_1)z_1)=\sum \limits_{p=1}^{\infty}\frac{1}{p!}
g_p(t_1, \dots ,t_1)z_1^p.  \nonumber
\end{equation}

The above expression should hold for arbitrary $z_1$.
Therefore expanding the logarithm in series, and equating the coefficients by
the same powers of $z_1$ on the both sides, one obtains the identity
\begin{equation}\label{E:CorrFuncCoinc}
g_p(t_1, \dots, t_1)=(-1)^{p-1}(p-1)! n_1^p(t_1).
\end{equation}
Eq.(\ref{E:CorrFuncCoinc}) is exact for all arguments coinciding.
Eq.(\ref{E:CorrFuncApp}) gives a correction to it, when the arguments differ.

Substitution of Eq.(\ref{E:CorrFuncApp}) into Eq.(\ref{E:Sexact}) delivers then the
\textit{Stratonovich approximation} for $\mathcal{F}(T)$ in form Eq.(\ref{E:GenForm}), 
now with the time-dependent escape rate being
\begin{equation}
S(T)=-\int \limits_0^Tn_1(t)\frac{\ln \left[
1-\int_0^T R(t,t^{\prime })n_{1}(t^{\prime })dt^{\prime }\right] }
{\int_0^T R(t,t^{\prime })n_{1}(t^{\prime })dt^{\prime }}dt.
\label{E:Straton}
\end{equation}

Let us now discuss the domains of applicability for these approximations.
The Hertz approximation Eq.(\ref{E:Hertz}) holds if all correlations decay
considerably within the typical time interval between upcrossings $T_R$.
The decay of correlations is described by the relaxation time
$t_{rel} =2/\gamma$ of the process. Therefore, the Hertz approximation
holds for $t_{rel} \ll T_R$.

The Stratonovich approximation is applicable when the argument of the
logarithm in Eq.(\ref{E:Straton}) is positive,
$1-\int_0^T [ n_1(t^{\prime}) +
n_1(t)^{-1} n_2(t^{\prime},t) ]dt^{\prime} >0$.
Using the fact that $n_2(t^{\prime},t) / n_1(t)$
tends to $n_1(t^{\prime})$ for $|t-t^{\prime}|>t_{rel}$ and tends to zero
for $|t-t^{\prime}| \rightarrow 0$ we get as a rough estimate for the validity region of
Eq.(\ref{E:Straton}) $t_{rel} < T_R$.

\begin{figure}[tbh]
\begin{center}
\epsfig{file=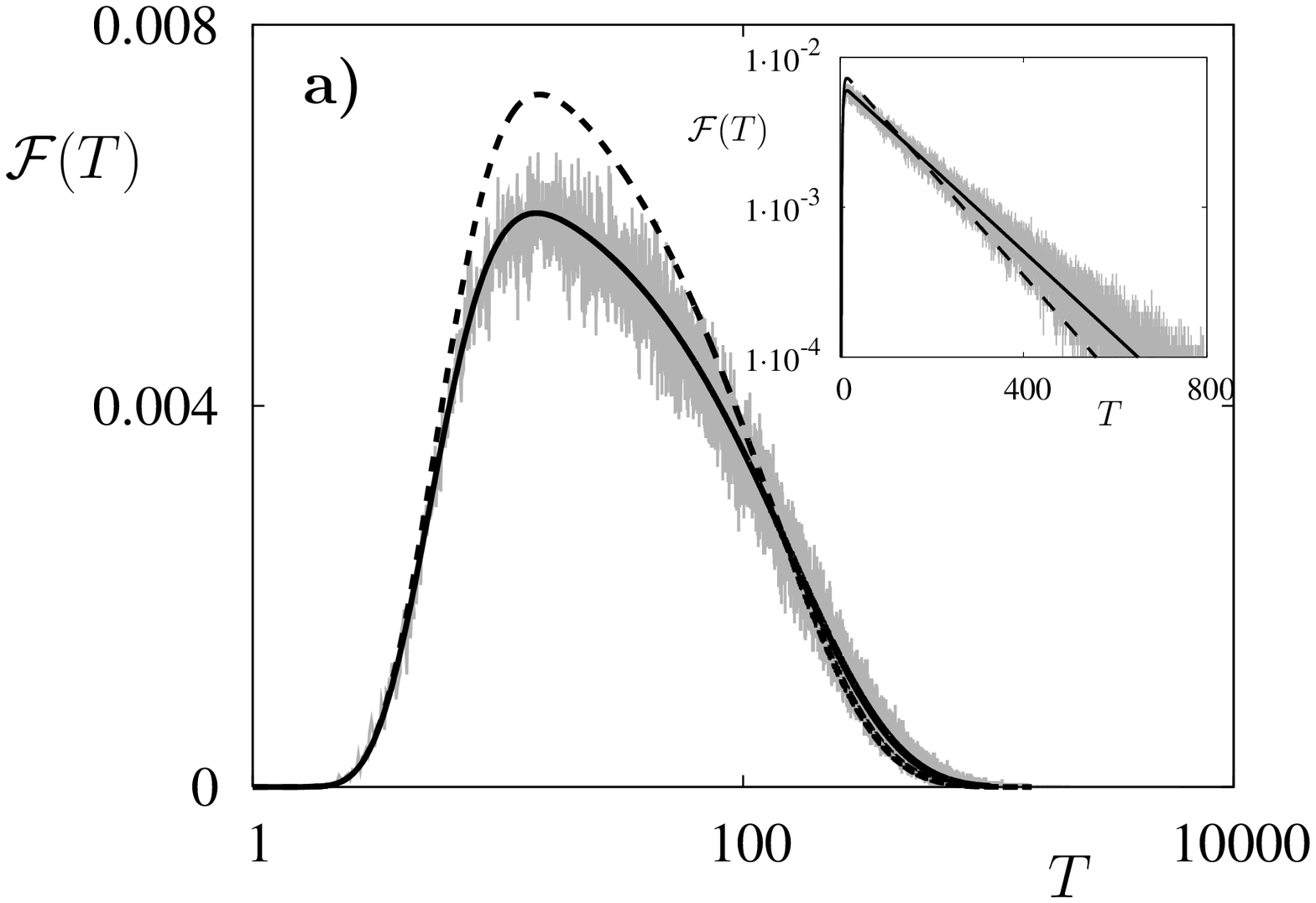,width=7.5cm}
\epsfig{file=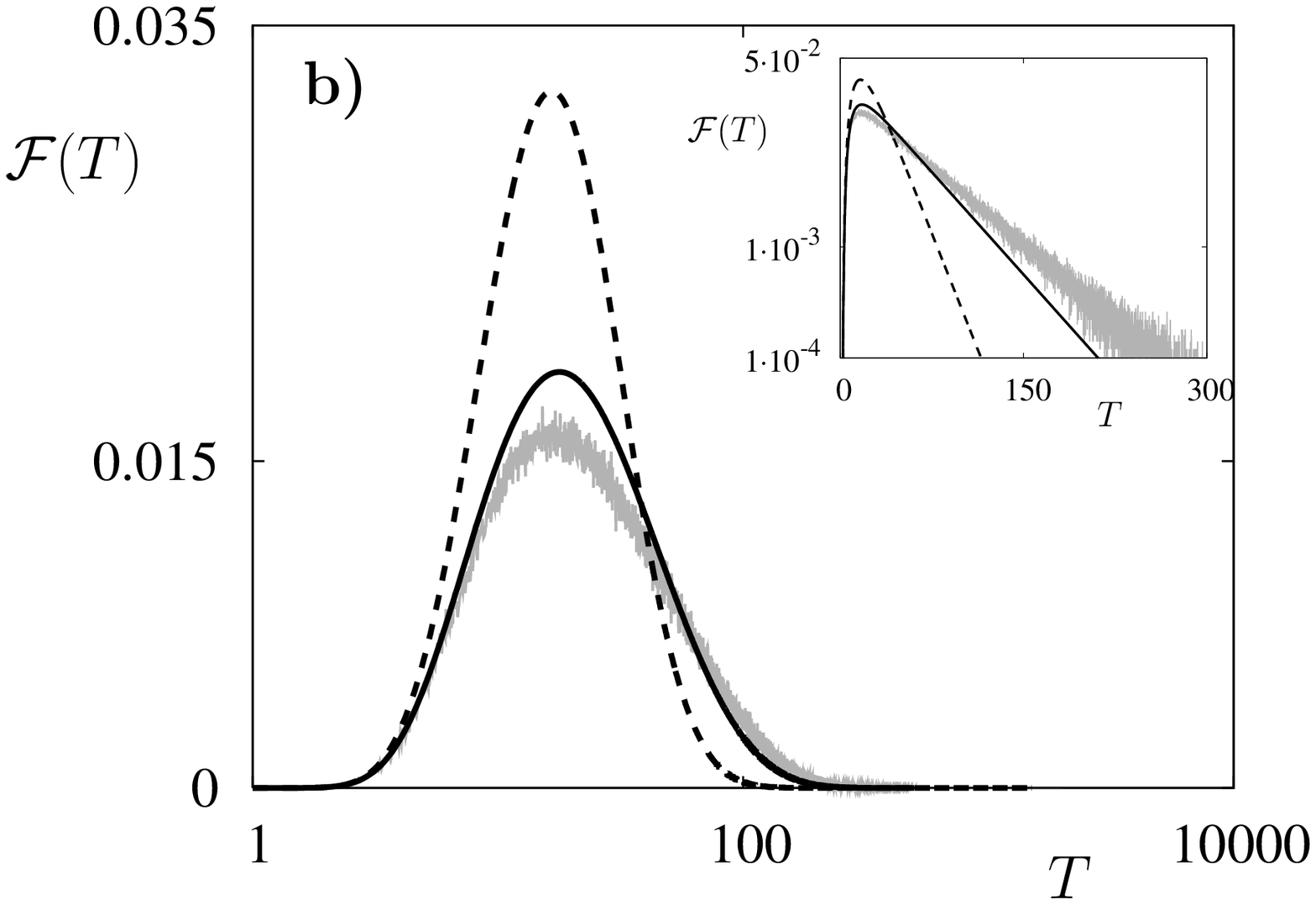,width=7.5cm}
\end{center}
\caption{Same as in Fig. \ref{F:FPTunderdamped} however
for the case of stronger friction. The parameters are
a) $\gamma = 3.0, D=0.5, \omega_0/\gamma=0.33$,
b) $\gamma=10.0, D=5.5, \omega_0/\gamma=0.1$ and other parameters
as in Fig. \ref{F:FPTunderdamped}.}
\label{F:FPToverdamped}
\end{figure}

Let us now turn to the results for the harmonic oscillator
Eq.(\ref{E:HarmOscil}) with white noise driving.
In Figs.\ref{F:FPTunderdamped} and \ref{F:FPToverdamped}
the FPT probability density obtained from simulations is depicted with a
gray line, the Hertz approximation Eq.(\ref{E:Hertz})
with a black dashed line, and the Stratonovich approximation
Eq.(\ref{E:Straton}) with a black solid line.

In Fig.\ref{F:FPTunderdamped}(a) the parameters are chosen to be
$\gamma=0.8,D=0.1$, corresponding to moderate friction and moderate
noise intensity. For given parameter values $t_{rel}=2.5$ and
$T_R=343$, so that $t_{rel} \ll T_R$, both Hertz and Stratonovich
approximations hold and reproduce well the FPT density in the whole
time domain.

In the case of moderate friction and stronger noise the upcrossings become
more frequent and $T_R$ decreases. The FPT changes its form to practically
monomodal. An example is given in Fig.\ref{F:FPTunderdamped}(b) with
$\gamma=0.8,D=0.44$ which correspond to $t_{rel}=2.5$ and $T_R=15.6$. The
Stratonovich approximation complies very well with simulations, while
the Hertz approximation fails to reproduce the details of the distribution:
It underestimates $\mathcal{F}(T)$ on short times, and shows slower
exponential decay in the tail than the one observed in simulations (see the
inset).

Finally, for small friction and week noise the upcrossings are rare, but
the relaxation time is large. The FPT probability density exhibits multiple
decaying peaks. In Fig.\ref{F:FPTunderdamped}(c) $\gamma=0.08,D=0.01$
corresponding to $t_{rel}=25, T_R=343$. Again, the Stratonovich
approximation performs well, while the Hertz approximation underestimates
the first peak, overestimates all further peaks and decays in the tail
faster than the simulated FPT density.

For the large $T$, $\mathcal{F}(T)$ decays exponentially, $\mathcal{F}(T) \propto
\exp(- \kappa T)$.  The decrement of this decay is obtained from long
time asymptotic: $\kappa T=\lim_{T \rightarrow \infty} S(T)$. Thus,
in the Hertz approximation Eq.(\ref{E:Hertz}) one gets
$\kappa_H=\lim_{T \rightarrow \infty} (1/T) \int_0^T n_1(t) dt =
n_0T/T=n_0 $.  The behavior in the Stratonovich approximation
Eq.(\ref{E:Straton}) is determined by $\lim_{t,t^{\prime} \rightarrow
  \infty} \int_0^{T} R(t,t^{\prime})n_1(t^{\prime})dt^{\prime} \approx
n_0 \tau_{cor}$ with $\tau_{cor}$ given by $\tau_{cor}= \lim
\limits_{t \rightarrow \infty} \int_0^{\infty}R(t,t^{\prime})
dt^{\prime}$. Note that $\tau_{cor}$ is not necessary positive
because of oscillating correlation coefficient.
Inserting this expression into Eq.(\ref{E:Straton}) and expanding the
logarithm up to the second term we get $\kappa_S =
n_0(1+\frac{1}{2}n_0\tau_{cor})$ providing the second order correction
to $\kappa_H$.  The value of $\tau_{cor}$ for the
parameter set as in Fig.\ref{F:FPTunderdamped}(a) is
$\tau_{cor}=-2.4$, for parameters as in
Fig.\ref{F:FPTunderdamped}(b) $\tau_{cor}=5.09$, and for parameters as in
Fig. \ref{F:FPTunderdamped}(c) $\tau_{cor}=-431.99$. The long time
asymptotic obtained with these $\tau_{cor}$ values reproduce fairly
well the decay patterns found numerically.

In the overdamped regime ($\gamma>2\omega_0$) the condition
$t_{rel}<T_R$ is always fulfilled. Nevertheless
the validity region of our approximations is limited.
With increasing friction the process $x(t)$ approaches to the markovian one
(it is markovian in the overdamped limit $\omega_0/\gamma \ll 1$).
For such processes the pattern of upcrossings is not homogeneous,
but shows rather well separated clusters of upcrossings \cite{S1967}.
Essentially in the markovian limit the property, that upcrossings form
a system of \textit{nonapproaching} random points is violated.
The upcrossings within a single cluster are not independent
even if their mean density $n_0$ is low, so that the quality of approximations
decreases. This fact is illustrated in Fig.\ref{F:FPToverdamped}.
In the overdamped regime the correlation functions decay monotonously,
and the FPT densities are always monomodal.
The parameters in Fig.\ref{F:FPToverdamped}(a) are
$\gamma=3.0,D=5.5$, so that $\omega_0/\gamma=0.33$.
Eqs.(\ref{E:GenForm}),(\ref{E:Straton}) continue to give a good approximation
for $\mathcal{F}(T)$, while the Hertz approximation becomes inaccurate.
Further increase in friction,
for example $\gamma=10.0,D=5.5$ as in Fig.\ref{F:FPToverdamped}(b)
corresponding to $\omega_0/\gamma = 0.1$, makes the process to approach the
Markovian limit. The Stratonovich approximation starts to be inaccurate,
and the Hertz approximation fails.


\section{Truncation versus decoupling approximations.}

In the two previous sections we have seen, that truncation
approximations reproduce the FPT density on short time scales.
In contrast, the decoupling approximations
reproduce FPT densities in the whole time domain and posses
the right normalization. At the first glance, it may seem
that the decoupling approximations excel the direct truncations
and should be preferably used in applications. However, it
depends on the problem one has to solve, and sometimes
the truncations turn out to be useful.

\begin{figure}[tbh]
\begin{center}
\epsfig{file=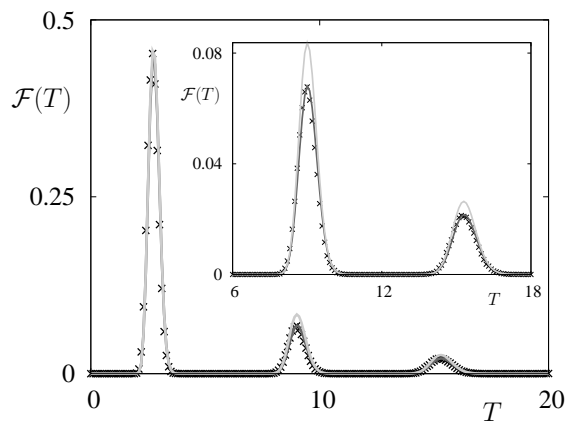,angle=90,width=7.5cm}
\end{center}
\caption{FPT density for harmonic oscillator driven by
white Gaussian noise, $\omega_0=1, \gamma=0.08, D=0.01,
x_0=-1, v_0=0, x_b=1$. Simulation results are shown
with crosses, the 3 terms approximation with black
line, and the Stratonovich approximation with gray
line. The inset shows the magnification of a part of the curve.}
\label{F:Str3terms}
\end{figure}

One such situation was already mentioned. If the noise
intensity is high or the barrier value is low, then the upcrossings
occur frequently, and the decoupling approximations can not be applied.
For example, for parameter values as in Fig.\ref{F:FPTG0.01D0.02},
the relaxation time is $t_{rel}=200$, and the mean interval
between upcrossings $T_R=8.07$. Thus $t_{rel}>T_R$, and both
Hertz and Stratonovich approximations fail.
Nevertheless, the most part of the FPT density is concentrated in
the first few peaks in this case, and is well reproduced
by the truncation approximations, as shown in
Figs.\ref{F:FPTG0.01D0.02} and \ref{F:FPTG0.01D0.02LSG}.

One can also be interested in a very accurate approximation for the
FPT density on short times. The truncations deliver better results
on short times than the decoupling approximations. For example,
in Fig.\ref{F:Str3terms} the simulation results are compared with
the 3 terms truncation and the Stratonovich approximation for the
same parameter values as in Fig.\ref{F:FPTunderdamped}(c).
Both approximation reproduce very accurate the first peak in
the FPT density. However, the 3 terms truncation is much more accurate
in estimation of the second and third peaks (see the inset
in Fig.\ref{F:Str3terms}).

This can be easily understood. The Stratonovich approximation
takes $n_1(t_1)$ and $n_2(t_2,t_1)$
exactly, and approximates all higher order densities through
these two. On times, when the first peak occurs, $n_2(t_2,t_1)$
is negligibly small. Then from Eqs.(\ref{E:CorrFuncApp}),(\ref{E:CorCoef})
we obtain $g_p(t_p,\dots, t_1)\approx (-1)^{p-1}(p-1)!n_1(t_p)\dots n_1(t_1)$.
Substitution of this expression into Eq.(\ref{E:CorrFuncGeneral})
gives then $n_p(t_p,\dots, t_1) \approx 0$ for $p>1$.
Thus on these times the Stratonovich approximation just coincides with
the 1 term truncation, and so reproduces the first peak very accurately.
On times, when the second peak in the FPT density occurs,
$n_2(t_2,t_1)$ is significantly different from zero. Hence all
approximated $n_p(t_p, \dots ,t_1), p>2$ turn out to be nonvanishing
as well, while the real values for these functions are
negligibly small on these times. Thus the accuracy of the Stratonovich
approximation decreases in the second peak.
From analogous reasoning it becomes clear, that the Hertz
approximation is already inaccurate in estimation of the first peak.


\section{Harmonic oscillator driven by colored noise}

Expressions Eq.(\ref{E:FPTgeneral}) and Eqs.(\ref{E:GenForm}),(\ref{E:Straton})
can be used to obtain the FPT density for a random process $x(t)$, if the
joint probability densities of $x$ and its velocity $v$ Eq.(\ref{E:RiceJoint})
exist. Thus it is necessary that the process $x(t)$ is continuous and differentiable
at any time, but there are no further restrictions on dimension and form of the
system. The truncation and decoupling approximations
deliver good results for the FPT density in their validity regions
independently of the character of the noisy drive. In particular the case of correlated
input signals (colored noise driving) is of importance
in neuroscience. For example, synaptic filtering of the input spike train
may lead to an exponentially correlated input signal.

Therefore we consider as another example a resonate-and-fire neuron
Eq.(\ref{E:HarmOscil}) driven by the
Ornstein-Uhlenbeck noise. The correlation time of the process is $\tau$,
the variance $D/\tau$, and the correlation function
$\langle \eta(t) \eta(t+t^{\prime}) \rangle=(D/\tau)\exp(-t^{\prime}/ \tau)$.
In the limit $\tau \rightarrow 0$ the process tends to the white noise of
intensity $2D$. The correlation functions $r_{xx}(t),r_{xy}(t),r_{yx}(t),
r_{xv}(t)=r^{\prime}_{xx}(t),r_{vv}(t)=-r_{xx}^{\prime \prime}(t),
r_{yv}(t)=r_{yx}^{\prime}(t), r_{vy}(t)=-r_{xy}^{\prime}(t)$
can be obtained using Fourier transform as it was done for the white noise
case in Section \ref{S:model}.

Then $n_1(T)$ is again obtained analytically and has the form given by
Eq.(\ref{E:1termExpl}).
Now $\mu_{ij}=\mu_{ji}$ are elements of the inverse correlation matrix
$(\hat{C}_5)^{-1}$, $q_i$ are components of vector
$\vec{Q}=(x(T),v(T),x_0(0),v_0(0),\eta_0(0))$. The factor $\alpha$ is defined in the
same way  as it was done in Section \ref{S:model}.

For simplicity, we assumed sharp initial conditions for the noise variable,
i.e. $\eta$ is reset after every spike to a fixed value $\eta_0$.
The alternative assumption, that the neuron variables $x,v$
are reset to their initial values once $x(t)$ reaches the thereshold
\textit{without} resetting $\eta(t)$, might be more realistic \cite{B2004}.
In this case all probability densities should be averaged with respect to the
stationary density of noise values \textit{upon firing}. However,
as an example, we confine ourselves to consideration
of sharp initial conditions for the noise: $P(\eta,t=0)=\delta(\eta-\eta_0)$.

\begin{figure}[tbh]
\begin{center}
\epsfig{file=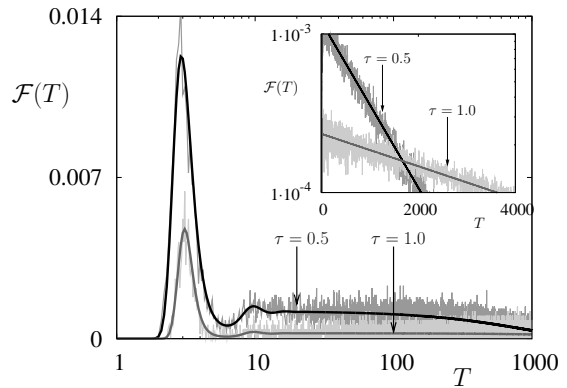,width=7.5cm}
\end{center}
\caption{Simulation results and the Hertz approximation for the FPT
probability density for harmonic oscillator driven by Ornstein-Uhlenbeck
noise with correlation times $\tau=0.5$ (upper curves) and $\tau=1.0$ (lower
curves) and other parameters as in Fig.\ref{F:FPTunderdamped}(a). Note the
logarithmic scale in $T$. The inset shows the same curves on the logarithmic
scale in $ {\cal F}(T)$.}
\label{F:FPTOU}
\end{figure}

In Fig.\ref{F:FPTOU} we show simulated FPT probability density
and the Hertz approximation for the harmonic oscillator driven by
the Ornstein-Uhlenbeck noise. We choose two different values of the correlation
time: $\tau=0.5$ and $\tau=1.0$. The reset value for the noise is $\eta_0=0$,
and other parameters are as in Fig.\ref{F:FPTunderdamped}(a).
The noise intensity decreases with increasing $\tau$, hence
the mean FPT growth. Though $\mathcal{F}(T)$ preserves its structure as a whole,
for larger values of $\tau$ the height of the main peak decreases and the weight
of the exponential tail growth. For given values of the correlation time
$T_R=829.2$ and $T_R=4247.1$ respectively, what significantly exceeds $t_{rel}$.
Thus in both cases the Hertz approximation is absolutely sufficient.


\section{Summary}

Motivated by studies of resonant neurons, we consider the first
passage time problem for systems with subthreshold oscillations
and nonnegligible relaxation times after a reset.
The joint densities of multiple upcrossings for such a process $x(t)$
can be obtained in the case of differentiable trajectories.
The FPT density for $x(t)$ is expressed in terms of an infinite series of
multiple integrals over all joint densities of upcrossings, or
equivalently, in terms of the cumulant functions.

We consider two types of approximations for this infinite series.
The truncation approximations include the first few terms of the series
calculated exactly. They reproduce well the FPT density on short and
intermidiate times and can be used, when the most part of the
FPT probability is concentrated in the first few peaks, i.e. when the barrier
value is low or the noise is strong.

The decoupling approximations can be derived for the case of weakly correlated
upcrossings. The higher order cumulant functions are expressed through the lower
order ones, and then infinitely many terms sum up to the closed
expression for $\mathcal{F}(T)$.  The Hertz approximation (the one neglecting all correlations
between upcrossings) is absolutely sufficient for the case of moderate friction
and moderate noise intensity. The Stratonovich approximation (approximating
the higher order cumulant functions through the first and the second ones)
performs even better and does not loose the accuracy for high noise intensities
or in the slightly overdamped regime.

We illustrate our results by the noise driven harmonic oscillator,
with the threshold value at $x_b$ and reset to sharp initial conditions, i.e.
the resonate-and-fire model of a neuron. The validity regions of
the approximations cover all different types of subthreshold dynamics.
Thus the approximations reproduce all
qualitatively different structures of the FPT densities: from monomodal
through bimodal to multimodal densities with several decaying peaks.
The approximations hold for systems of whatever dimension.
We illustrate this by the harmonic oscillator driven by the
Ornstein-Uhlenbeck noise.

Though we applied the theory to the harmonic oscillator
(resonate-and-fire model), the linearity of the system is in general not
required. The joint distributions of $x(t)$ and its velocity should exist,
i.e. $x(t)$ should be differentiable in time. No further restrictions
on the form and dimension of the system are implied.

\section*{Acknowledgments}

We acknowledge financial support from the DFG through Graduierten Kolleg 268
and Sfb 555.

\end{document}